\documentstyle[sprocl]{article}

\def\Journal#1#2#3#4{{#1,} {{\bf #2},} #3 #4}

\def\be{\begin{equation}}
\def\ee{\end{equation}}
\def\bea{\begin{eqnarray}}
\def\eea{\end{eqnarray}}

\def\lessim{\lower0.6ex\hbox{$\,$\vbox{\offinterlineskip
\hbox{$<$}\vskip1pt\hbox{$\sim$}}$\,$}}

\def\grtsim{\lower0.6ex\hbox{$\,$\vbox{\offinterlineskip
\hbox{$>$}\vskip1pt\hbox{$\sim$}}$\,$}}

\font\tenbb=msym10
\font\sevenbb=msym7
\font\fivebb=msym5
\newfam\bbfam
\textfont\bbfam=\tenbb \scriptfont\bbfam=\sevenbb
\scriptscriptfont\bbfam=\fivebb
\def\bb{\fam\bbfam}

\def\Nb{{\bb N}}

\def\buildrel#1\over#2{\mathrel{
\mathop{\kern 0pt#2}\limits^{#1}}}

\def\build#1_#2^#3{\mathrel{
\mathop{\kern 0pt#1}\limits_{#2}^{#3}}}

\begin{document}

\title{THEORETICAL ASPECTS OF GRAVITATIONAL RADIATION\footnote{Talk
given at GR14 (Firenze, Italy, 6-12 August 1995); to appear in the
proceedings edited by M. Francaviglia et al (World Scientific, Singapore).}}

\author{ T. DAMOUR }

\address{Institut des Hautes Etudes Scientifiques,\\ 
91440 Bures-sur-Yvette, France\\
{\rm and}\\
D\'epartement d'Astrophysique Relativiste et de Cosmologie,\\
Observatoire de Paris, Centre National de la Recherche Scientifique,\\ 
92195 Meudon, France}

\maketitle\abstracts{
A central problem in gravitational wave research is the {\it
generation problem}, i.e., the problem of relating the outgoing
gravitational wave field to the structure and motion of the material
source. This problem has become, in recent years, of increased
interest in view of the development of a worldwide network of
gravitational wave detectors. We review recent progress in
{\it analytical} methods of tackling the gravitational wave
generation problem. In particular, we describe recent work in an
approach which consists of matching a post-Newtonian expansion of the
metric near the material source with a multipolar-post-Minkowskian
expansion of the external metric. The results of such analytical methods
are important notably for providing accurate theoretical predictions for
the most promising targets of the LIGO/VIRGO interferometric
network: the ``chirp'' gravitational waveforms emitted during the
radiation-reaction-driven inspiral of binary systems of compact
objects (neutron stars or black holes).
}

\section{Introduction}

I wish to dedicate this talk to Henri Poincar\'e who introduced (several
aspects of) the concept of gravitational wave (``onde gravifique'')
ninety years ago. Indeed, in his two seminal papers of June~\cite{P5}
and July~\cite{P6} 1905 (the first one of which preceded Einstein's paper
on special relativity by one week), Poincar\'e not only introduced what
was to become later the basic mathematical structures of special
relativity (the Poincar\'e group and the ``Minkowski'' metric $(it)^2
+x^2 +y^2 +z^2$), but also pioneered the idea that one needed, for
consistency, a {\it relativistic} theory of gravitation. In his two
papers written in 1905, he defines a class of ``Poincar\'e invariant''
gravity theories and emphasizes that they predict that the gravitational
interaction propagates with the velocity of light~\cite{Pcit1}. In a later
work of 1908~\cite{P8}, he went as far as speaking of the emission of
gravitational waves (``onde d'acc\'el\'eration'') and of the associated
loss of energy of the emitting system. He even mentions that the main
observable effect of this dissipation of energy into gravitational waves
will be a secular acceleration of the mean motions of
planets~\cite{Pcit2}. It is interesting to note that the discovery of
binary pulsars~\cite{HT75}, and their subsequent continuous
observation~\cite{T94}, has allowed one precisely to verify the aspects
of gravitational waves discussed by Poincar\'e: propagation of the
gravitational interaction with finite velocity, and associated effect on
the orbital dynamics~\cite{DD}. This verification has been
observationally possible in two binary pulsar systems:
PSR1913+16~\cite{T94} and PSR1534+12~\cite{WT}.

The discovery of binary pulsars is important in three respects
for gravitational radiation research: (i) it establishes the reality of
gravitational radiation by verifying in a {\it direct} manner that
gravity propagates with the velocity of light between the companion and
the pulsar; (ii) it gives us our first tests of the {\it strong-field}
regime of gravity~\cite{DT92}, \cite{TWDW} thereby confirming the validity
of Einstein's theory in a regime so forth untested; (iii) it establishes
the existence of strong sources of gravitational waves, thereby providing
fascinating targets for the LIGO/VIRGO network of interferometric
detectors.

Indeed, the observation of the secular acceleration of the orbital mean
motions of the binary pulsars 1913+16 and 1534+12 proves that, in a few
hundred million years, these systems will have shrunk so much that they
will constitute an ``inspiralling'' binary system of neutron stars: i.e.
a very close system of two neutron stars, orbiting around each other at a
very fast and accelerated pace, the orbital frequency increasing from, say,
$\sim 10$ Hz to $\sim 1000$ Hz in about 20 minutes. Then, when the two
stars get near each other they start coalescing together to form only one
central object. In the last minutes of the inspiralling motion such
systems emit rather strong gravitational waves. The first person to
conceive of such systems, and to realize they provided superb targets for
gravitational wave detectors, was Dyson~\cite{D63}, in a prescient paper
written years before the discovery of pulsars established the existence
of neutron stars. The characteristics of inspiralling and coalescing
binaries as gravitational wave sources have then been explored by several
authors~\cite{FB67}, \cite{CE77}, \cite{T87}, \cite{S86}. The rate of
occurrence of such events seem to be high enough to furnish a regular
($\sim$ monthly) source of signals for the LIGO/VIRGO network~\cite{P91},
\cite{NPS91}. The information content of inspiralling events is of
excellent quality: their detection (if done with a suitably high signal
to noise ratio) should allow one: (i) to measure directly cosmological
distances and thereby to have a clean access to the cosmological
parameters $H_0$ and $q_0$~\cite{S86}, \cite{KS87}, \cite{C93},
\cite{M93}, \cite{CF93}; (ii) to test the nonlinear structure of
radiative gravity~\cite{BS95}; (iii) to perform new tests of the
existence of a scalar component to gravity~\cite{W94}; (iv) to probe
black hole physics~\cite{RFT95}. In view of the importance of
inspiralling events, it is crucial to dispose of an accurate theoretical
model of the corresponding gravitational wave signals. The aim of the
present contribution is to review recent progress in {\it analytical
methods} of tackling the generation of gravitational waves, and their
application to inspiralling binaries.

\section{Analytical formalisms for treating the generation of
gravitational waves}

One can distinguish three basic questions in gravitational radiation
theory:

\noindent -- Question 1 (``asymptotic problem''): What is the asymptotic
behaviour, appropriate to isolated systems and consistent with Einstein's
field equations, of radiative gravitational fields far from their sources?

\noindent -- Question 2 (``generation problem''): What is the link between
the preceding asymptotic behaviour and the structure and motion of the
sources that generate the gravitational radiation?

\noindent -- Question 3 (``radiation reaction problem''): What is
the back-reaction of the emission of gravitational radiation on the source?

The standard answer to Question 1 is given by the Bondi-Sachs-Penrose
description of radiative, asymptotically flat spacetimes, with a
sufficiently smooth fall off at ${\cal I}^+$ and the exclusion of ingoing
waves on ${\cal I}^-$. However, this answer is still a conjecture. Let us
note, in this respect, that (i) the estimates used in the global theorem
of Christodoulou and Klainerman~\cite{CK} are not strong enough to
establish the standardly assumed peeling at ${\cal I}^+$; and (ii) some
perturbation calculations suggest a violation of peeling in scattering
problems~\cite{D86} (where it is found that ${\cal I}^+$ cannot be $C^3$).

Questions 2 and 3 have standard answers (discussed in textbooks) only at
the lowest approximation. These standard answers go by the (ambiguous)
name of ``quadrupole formulas''. Actually, one should carefully
distinguish: the ``far-field quadrupole formula'', the ``energy-loss
quadrupole formula'', the ``radiation-reaction quadrupolar force'',
etc$\ldots$ (see, e.g., \cite{D87} for a discussion). In any case, these
standard answers are insufficiently accurate to give mathematical models
of inspiralling signals adequate for high-precision observations. Indeed,
during the final stages of the inspiralling motion the orbital velocities
become rather high $(v/c \lessim 0.3)$ and necessitate the consideration
of many corrections to the leading ``quadrupole'' result.

Several methods have been proposed for going beyond the lowest-order
results. For instance, some years ago Epstein, Wagoner and
Thorne~\cite{EW75}, \cite{T80} (in an attempt to generalize the
Landau-Lifshitz-type derivation of the standard far-field quadrupole
formula) introduced a post-Newtonian extension of the quadrupole
formalism. Though their formalism is marred by some mathematical
difficulties (divergent integrals), it was used to derive ${\cal O} (v^2
/c^2)$ corrections to the quadrupole formula for binary
systems~\cite{WW76}, and has been recently used~\cite{WW} to go to higher
orders in $v/c$. A basic problem of the
Landau-Lifshitz-Epstein-Wagoner-Thorne approach is the lack of a clear
separation between the near zone and the wave zone. The combined use of
an ``effective'' stress-energy tensor for the gravitational field (with
non-compact support) and of formal post-Newtonian expansions quickly
leads to the appearance of divergent integrals. By contrast, Blanchet,
Damour and Iyer~\cite{BD86}, \cite{B87}, \cite{BD88}, \cite{BD89},
\cite{DI91}, \cite{BD92}, \cite{B95a}, \cite{B95b} (building on the
Fock-type derivation of the quadrupole formula and on the
double-expansion method of Bonnor~\cite{B59}) introduced a new
gravitational-wave-generation formalism based on a clean separation
between near-zone and wave-zone effects. The Blanchet-Damour-Iyer
approach is mathematically well-defined and obtains corrections to the
leading quadrupolar formalism in the form of compact-support integrals.
[In a recent development of this formalism, Blanchet~\cite{B95a} found it
convenient to obtain the corrections in the form of (well-defined)
analytically-continued integrals which are (formally) equivalent to
compact-support integrals.] The BDI scheme has a ``modular structure'':
the final results are obtained by combining an ``external zone module''
(in which the external, vacuum metric is expanded as a
multipolar-post-Minkowskian double series) with a ``near zone module''
(based on a more traditional post-Newtonian-type expansion). When dealing
with strongly self-gravitating material sources (such as neutron stars,
or black holes) one must also use a ``compact body module''~\cite{D83}.
After elimination of the various mathematical intermediaries appearing in
the formalism (such as the ``algorithmic'' multipole moments $M_L$ and
$S_L$), the basic structure of the final results of the BDI formalism is
the following: the (directly observable) ``radiative'' multipole moments
$U_L$ and $V_L$~\cite{Not}, parametrizing the angular dependence of the
asymptotic gravitational wave amplitude $h_{ij}^{TT}
(T,R,\theta,\varphi)$, are given in terms of the ``source'' multipole
moments $I_L$ and $J_L$ as a series of terms of increasing nonlinearity:

$U_L (t) = d^{\ell} I_L (t) / dt^{\ell} + {\cal F}_L [I(t'),J(t')]$,

$V_L (t) = d^{\ell} J_L(t) / dt^{\ell} + {\cal G}_L [I(t'),J(t')]$,

\noindent where ${\cal F}_L$ and ${\cal G}_L$ are multilinear (successively
quadratic, cubic, etc$\ldots$) {\it retarded} functionals of the full
past behaviour of the source moments $I_{L'} (t')$, $J_{L'} (t')$ (for
$t'\leq t$). As for the ``source'' moments $I_L (t)$, $J_L (t)$ they are
(approximately) defined in the formalism as some {\it instantaneous}
(nonlinear) functionals of the source variables $T^{\mu \nu} (\hbox{\bf x}
,t)$ [or $m_A$, $\hbox{\bf z}_A (t)$ in the case of spherical compact
bodies].

For instance, the first two corrections (of order $v^2 /c^2$ and $v^3
/c^3$) to the radiative quadrupole moment are given by~\cite{BD89},
\cite{BD92}
\begin{equation}
U_{ij} (t) = \frac{d^2 I_{ij} (t)}{dt^2} + \frac{2GM}{c^3}
\int_{0}^{\infty} d\tau \left( \ln \frac{\tau}{2b} + \frac{11}{12}
\right) \frac{d^4 I_{ij} (t-\tau)}{dt^4} +\cdots \label{one}
\end{equation}
with
\begin{equation}
\begin{array}{rcl}
I_L (t) &= &{\displaystyle \int} d^3 x \widehat{x}^L \sigma (t,\hbox{\bf
x}) + {\displaystyle \frac{1}{2(2\ell +3)c^2} \frac{d^2}{dt^2}}
{\displaystyle \int} d^3 x \widehat{x}^L \hbox{\bf x}^2 \sigma (t,\hbox{\bf
x}) \\ 
&& \\
&& -{\displaystyle \frac{4(2\ell +1)}{(\ell +1)(2\ell +3) c^2}
\frac{d}{dt} \int} d^3 x \widehat{x}^{kL} \sigma_k (t,\hbox{\bf x})+\cdots
\\ & & \end{array} \label{two}
\end{equation}
where $\sigma \equiv (T^{00} + T^{kk})/c^2$, $\sigma_i \equiv T^{0i} /c$.
The integral appearing on the right of equation~\ref{one} represents the
effect of the backscattering of the gravitational waves on the
Schwarzschild-like curvature associated to the total mass $M$ of the source
(``tails'')~\cite{HR69}, \cite{BD92}. Beyond the terms written
in equation~\ref{one} there are many other nonlinear contributions (of
formal higher order in $v/c$). For instance at the quadratically nonlinear
order one has a contribution to $U_L (t)$ depending upon the gravitational
wave flux emitted in the past,
\begin{equation}
\frac{8\pi c^{\ell -2} \ell!}{(\ell +1)(\ell +2)} \int_{-\infty}^{t} dt'
\left( \frac{dE^{GW} (t')}{dt' d\Omega} \right)_L \, , \label{three}
\end{equation}
which has been discussed (in different guises, and under different names)
by several authors~\cite{P83}, \cite{BDthese}, \cite{C91}, \cite{WW91},
\cite{T92}. For explicit applications of the presently discussed scheme,
see~\cite{BS89}, \cite{BS93}, \cite{KWW}, \cite{W93}, \cite{BS95},
\cite{DI94}, \cite{D95}, \cite{BDI95}, \cite{K95}. Besides ``hereditary''
effects in the wave zone (such as the integral contributions in
equations~\ref{one} and~\ref{three}), it has also been possible to
investigate the leading hereditary effects appearing in the near-zone
field: they enter at the fourth post-Newtonian level, i.e. $(v/c)^8$
beyond the Newtonian approximation, and correspond to ``tail''
modifications of the ${\cal O} (v^5 /c^5)$ Burke-Thorne radiation
reaction potential~\cite{BD88}.

\section{Inspiralling compact binaries}

The accurate mathematical modelling of the gravitational wave signals
emitted by inspiralling compact binaries needs two (related) inputs: (i)
a solution of the ``generation problem''; and (ii) a solution of the
``radiation-reaction problem'' adequate for treating compact objects. The
schemes discussed in the previous section were primarily aimed at solving
the generation problem, i.e. at giving $h_{ij}^{TT}$ as a retarded
functional of the structure and motion of the source. However, to have an
explicit representation of $h_{ij}^{TT}$ as a function of time, one needs
to know the time evolution of the source, i.e. to solve the problem of
motion, including radiation-reaction effects. Actually, it has been
recently emphasized~\cite{C93} that the radiation-reaction part of the
problem was the most crucial one in that it determined the time evolution
of the {\it phase} of the gravitational wave signal (which follows,
modulo a factor two for circular orbits, the orbital phase). [We work here
within the ``restricted waveform'' approximation, i.e. we focus on the main
Fourier component of the signal.] Indeed, an accurate modelling of the
(radiation-reaction-driven) phase $\phi_{GW} (t)$ of the gravitational wave
is essential for a successful detection based on correlating the observed,
noisy $h^{\rm obs} (t)$ with some theoretical template $h^{\rm theory} (t) =
a_{GW} (t) \cos \phi_{GW} (t)$~\cite{C93}, \cite{FC93}, \cite{CF94} (more so
than the modelling of the evolution of the amplitude $a_{GW} (t)$). The
results discussed in the previous section are certainly accurate enough for
predicting $a_{GW} (t)$. The situation for what concerns the
radiation-reaction driven phase $\phi_{GW} (t)$ is less satisfactory.
Indeed, the only {\it complete} results available for the equations of
motion of a compact binary are the $(v/c)^5$-accurate equations of
motion~\cite{DD}. Beyond this level, only {\it partial} results are known:
namely the ${\cal O} ((v/c)^7)$ radiation-reaction terms~\cite{IW93},
\cite{B95c}, and the ${\cal O} ((v/c)^8)$ hereditary contribution to the
radiation-reaction~\cite{BD88}. The (expected) ``conservative''
contributions (of order $(v/c)^6 + (v/c)^8 +\cdots$) to the equations of
motion are unknown, as well as the higher-order contributions to radiation
reaction. The only way one can presently deal with this problem is to {\it
heuristically} rely on a (naive) energy-balance argument to relate the loss
of {\it mechanical} energy of the binary system to the asymptotic flux of
gravitational waves. In technical terms, one writes \begin{equation}
\begin{array}{rcl}
& {\displaystyle \frac{dE^{\rm mechanical}}{dt}} = -{\displaystyle
\frac{dE^{GW \, {\rm flux}}}{dt}} = {\displaystyle \sum_{\ell =2}^{\infty}
\frac{G}{c^{2\ell +1}}} \\ 
& \\
& \times \left[ {\displaystyle \frac{(\ell +1)(\ell +2)}{(\ell -1) \ell
\ell ! (2\ell +1)!!} \left( \frac{dU_L}{dt} \right)^2} + {\displaystyle
\frac{4\ell (\ell +2)}{c^2 (\ell -1)(\ell +1)! (2\ell +1)!!} \left(
\frac{dV_L}{dt} \right)^2} \right] \, , \\ 
& 
\end{array} \label{four} 
\end{equation}
in the right-hand side of which one inserts the best available results
(from solving the generation problem) on the radiative multipole moments
generated by a compact binary.

The highest-accuracy results obtained along these lines have been derived
in the limiting case of a very small test mass orbiting a heavy central
mass modeled as a Schwarzschild (or Kerr) black hole. Indeed, in such a
case one can treat the generation problem as a linear perturbation of
Schwarzschild or Kerr. The latter pertubation problem can, thanks to the
work of many people (Regge, Wheeler, Zerilli, $\ldots$, Teukolsky,
$\ldots$, Sasaki, Nakamura, $\ldots$, Chandrasekhar, $\ldots$), be reduced
to integrating some linear ordinary differential equations for the radial
dependence. This integration can be done numerically~\cite{P*},
\cite{TN94}. Moreover, the post-Newtonian expansion (in powers of $v/c$) of
the generated gravitational wave amplitudes can be derived
analytically~\cite{P*}, \cite{S94}, \cite{TS94}. The results of this
approach are important testbeds for the full problem and can help us in
posing the important questions (such as: ``how fast does the
post-Newtonian expansion converge?''~\cite{C93}), but they fall short of
providing us with the answers we really care about. Indeed, if we
introduce the dimensionless parameter measuring the deviation from the
test mass limit, \begin{equation}
\nu \equiv \eta \equiv \frac{m_1 m_2}{(m_1 +m_2)^2} \label{five}
\end{equation}
(where $m_1$ and $m_2$ are the masses of the members of an inspiralling
binary), we expect that many of the systems that LIGO/VIRGO will detect
will be made of two nearly equal neutron star masses: $m_1 \approx m_2
\approx 1.4 M_{\odot}$. This corresponds to the maximum possible
deviation from the test-mass limit ($\nu =\frac{1}{4}$ instead of $\nu
\ll 1$) for which the black-hole perturbation results become unreliable.
This is why we badly need the general-purpose analytical methods
discussed in the previous section. Only such methods can, at present,
deal with inspiralling binaries having comparable masses $m_1 \sim m_2$.
[Note, however, that black-hole perturbation methods are directly
relevant for dealing with some of the target sources of low-frequency space
interferometers such as LISA: e.g. the fall of a neutron star into a very
massive black hole.]

The main result one is interested in is an analytical expression giving
the time evolution of the gravitational wave phase, i.e. something we can
call the ``phasing formula'' of inspiralling binaries:
\begin{equation}
\phi^{GW} = 2\phi^{\rm ORBITAL} = F[t_{\oplus} ; p_i] \, , \label{six}
\end{equation}
where $t_{\oplus}$ is the (proper) time at the Earth laboratory recording
the continuous arrival of the (main) gravitational wave signal
\begin{equation}
h_{\oplus}^{GW} (t_{\oplus}) = a^{GW} (t_{\oplus}) \cos \phi^{GW}
(t_{\oplus}) \, , \label{seven} 
\end{equation}
and where $\{ p_i \}$ is a set of parameters carrying information about
the emitting binary system. It is interesting to note that the ``phasing
formula'' is nothing but a continuous analog of the discrete ``timing
formula'' which is basic to relativistic pulsar timing. Indeed, the
timing formula of binary pulsars~\cite{DD86}, \cite{DT92} can be written
as
\begin{equation}
\phi^{PSR}_N = F[t_N^{\oplus} ; p_i] \, , \label{eight}
\end{equation}
where $t_N^{\oplus}$ is the (proper) time of arrival at the Earth
observatory of the $N$th ($N\in \Nb$) pulse emitted when the rotational
phase of the spinning pulsar was $\phi_N^{PSR} \simeq 2\pi N
+\hbox{const}$. Here also $\{ p_i \}$ is a set of parameters carrying
information about the binary system, and the basic aim of pulsar timing
is to measure as many $p_i$'s as possible~\cite{DD86}, \cite{DT92}. The
analogy between equations~\ref{six} and \ref{eight} is clear: in one case
one is timing from Earth a continuous orbital phase, in the other case one
is timing a stroboscopic rotational phase.

Thanks to recent works which explicitly developed the analytical methods
discussed in the previous section to the $(v/c)^4$~\cite{BDIWW},
\cite{BDI95}, \cite{WW} and $(v/c)^5$~\cite{B95b} orders one knows the
phasing formula of inspiralling binaries (of arbitrary masses) to the
following accuracy: introducing the dimensionless time variable
\begin{equation}
\widehat t \equiv \frac{c^3 \nu}{5G(m_1 +m_2)} (t_c -t) \label{nine}
\end{equation}
(where $t_c$ denotes the coalescence time) we can write
\begin{equation}
\begin{array}{rcl}
\phi_c -\phi^{\rm ORBITAL} &= &{\displaystyle \frac{1}{\nu}}
\widehat{t}^{\frac{5}{8}} \Bigl\{ 1+A_2 (\nu) \widehat{t}^{-{\frac{2}{8}}}
+ A_3 (\nu)  \widehat{t}^{-{\frac{3}{8}}} + A_4 (\nu)
\widehat{t}^{-{\frac{4}{8}}} \\
&& \\
&& +  \, B_5 (\nu) \widehat{t}^{-{\frac{5}{8}}} \ln \widehat t 
 + {\cal O}
\left(\widehat{t}^{-{\frac{6}{8}}}\right) \Bigl\} \, , \\ & & 
\end{array} \label{ten} 
\end{equation}
where
\begin{equation}
A_2 (\nu) = \frac{5}{24} \left( \frac{743}{336} + \frac{11}{4} \nu \right)
\label{eleven} 
\end{equation}
($v^2 /c^2$ corrections: see~\cite{WW76}, \cite{BS89})
\begin{equation}
A_3 (\nu) =-\frac{3}{4} \pi \label{twelve} 
\end{equation}
($v^3 /c^3$ or ``tail'' corrections see~\cite{P*} for the $\nu =0$ limit,
and~\cite{BD92}, \cite{BS93}, \cite{W93} for $\nu \not = 0$)
\begin{equation}
A_4 (\nu) = \frac{5}{64} \left( \frac{1855099}{225792} +
\frac{56975}{4032} \nu + \frac{371}{32} \nu^2 \right) \label{thirteen} 
\end{equation}
($v^4 /c^4$ corrections: see~\cite{TS94} for the $\nu =0$ limit
and~\cite{BDIWW}, \cite{BDI95}, \cite{WW} for $\nu \not = 0$),
and 
\begin{equation}
B_5 (\nu) = -\pi \left( \frac{38645}{172032} + \frac{15}{2048} \nu \right)
\label{fourteen}
\end{equation}
($v^5 /c^5$ corrections, after the absorption of any $A_5 (\nu)$ in the
definition of $\phi_c$: see~\cite{TS94} for the $\nu =0$ limit
and~\cite{B95b} for $\nu \not = 0$). The numerical importance of
finite-mass-ratio effects ($\nu \not = 0$) is to be noted. In particular,
the equal mass case $\left( \nu =\frac{1}{4} \right)$ represents (with
respect to the test-mass limit) an increase of $A_2$ by 31\% and of $A_4$ by
52\%! (By contrast the corresponding change in $B_5$ is only 0.8\%.)

\section{Conclusions}

\noindent $\bullet$ A general comment on the present brief review of
analytical approaches to gravitational radiation is that it confirms the
perennial validity of a remark by Poincar\'e: to the effect that real
problems are never definitively solved, but only {\it more or less}
solved (``il y a seulement des probl\`emes {\it plus ou moins} 
r\'esolus~\cite{PSM}).

\noindent $\bullet$ Though I have insisted on the need of computing the
(hard to get) higher-order contributions to the phasing formula of
inspiralling binaries because of their importance for extracting the
maximum possible information from gravitational wave signals, it should
be stated that these corrections are (probably) not needed for searching
and discovering gravitational wave signals in the noise.

\noindent $\bullet$ The ultimate post-Newtonian accuracy which is really
needed in the phasing formula for an acceptably accurate determination of
the information-carrying parameters $\{ p_i \} = \{ t_c
,\nu^{\frac{3}{5}} (m_1 +m_2), \nu \}$ is still unclear at present
(see~\cite{C93}, \cite{TN94}, \cite{CF95}, \cite{P95}).

\noindent $\bullet$ I have considered above only the simplest case where
the members of an inspiralling system are slowly spinning. This is the
situation one can plausibly expect in most neutron star-neutron star
systems~\cite{BDIWW}. However, systems containing black holes (if they
exist in appreciable number) might contain fast spinning objects. See
e.g.~\cite{KWW}, \cite{K95} for spin-dependent effects.

\noindent $\bullet$ If very high post-Newtonian contributions to the
phasing formula are really needed, one might need to reconsider the
presently developed analytical approaches. It might, for instance, become
necessary to define them in a fully algorithmic manner allowing the use
of computer-based algebraic programmes, or it might become necessary to
match them to numerical relativity results (see e.g.~\cite{SNO92},
\cite{Wilson}). A better understanding of the mathematical nature of the
post-Newtonian expansion might also help.

\noindent $\bullet$ I anticipate that a serious obstacle to improving the
present $(v/c)^5$ accuracy of the phasing formula will come from the need
to extend the accuracy of the equations of motion of compact binaries
beyond the $G^3$ level treated in~\cite{D83}.

\noindent $\bullet$ The increasingly slow convergence of the
post-Newtonian series toward the end of the inspiralling stage points out
the importance of improving the sensitivity of the detectors on the
low-frequency side (say a few tens of Hz, corresponding to gravitational
waves emitted early on). I note, in this respect, that the VIRGO detector
puts a particular emphasis on improving its low-frequency sensitivity.

\bigskip

\noindent{\bf Acknowledgments}

\noindent I thank L. Blanchet for useful comments and for providing me with
equation \ref{fourteen}.


\begin{thebibliography}{99}
\bibitem{Not} Here $L=i_1 i_2 \ldots i_{\ell}$ denotes a multi-index with
$\ell$ spatial indices $i_p =1,2,3$. The tensors $M_L$, $S_L$, $I_L$,
$J_L$, $U_L$, $V_L$ are all symmetric and trace-free (STF). In
equation~\ref{two} $\widehat{x}^L$ denotes the STF part of the tensor $x^L
= x^{i_1} x^{i_2} \ldots x^{i_{\ell}}$.
\end{thebibliography}
\end{document}